# GLOBAL WARNING: CHALLENGES, THREATS AND OPPORTUNITIES FOR GROUND BEETLES (COLEOPTERA: CARABIDAE) IN HIGH ALTITUDE HABITATS

MAURO GOBBI

Section of Invertebrate Zoology and Hydrobiology, MUSE-Science Museum, Corso del Lavoro e della Scienza 3, I-38122 Trento (Italy), email: Mauro.Gobbi@muse.it; ORCID: https://orcid.org/0000-0002-1704-4857

*Abstract*

Aim of this paper is to provide the first comprehensive synthesis about ground beetle (Coleoptera: Carabidae) distribution in high altitude habitats. Specifically, the attention is focused on the species assemblages living on the most common ice-related mountain landforms (glaciers, debris-covered glaciers, glacier forelands and rock glaciers) and on the challenges, threats and opportunities carabids living in these habitats have to face in relation to the ongoing climate warming. The suggested role of the ice-related alpine landforms, as present climatic refugia for cold-adapted ground beetles, is discussed. Finally, the needs to develop a large-scale High-alpine Biodiversity Monitoring Program to describe how the current climate change is shaping the distribution of high altitude specialists, is highlighted.

**PREFACE**
Mountains cover over only a tenth of the continental surface of the Earth (excluding Antarctica) (KÖRNER *et al* 2017), but harbor one quarter of all terrestrial species (HOORN *et al* 2018). Therefore, it is quite easy to understand that a world without mountains would be a less diverse place in terms of both habitats and species.
What are the main factors driving such a high of biodiversity? Mountains i) have strong altitudinal gradients, and consequently thermal gradients, ii) are topographically very complex due to the mosaic of different landforms on different bedrocks, iii) can experience strong climatic diversity within a few kilometers (oceanic vs. continental climate) and iv) experienced a prolonged isolation of their peaks and valleys during the Quaternary climatic fluctuations. All these factors influenced and are still influencing the distribution and diversification of species. Famous naturalists and alpinists (e.g. Linnaeus, Humboldt and Whymper) had already recognized the importance of these factors for mountain biodiversity (HOORN *et al* 2018; MORET *et al* 2019).
Over fifty glacial and interglacial climate cycles occurred during the Quaternary period and played a major role in shaping the ranges of several species (HOORN *et al* 2018). Our current Holocene is an interglacial period (warm interval) that began about 11,000 yr BP. Starting at the end of the Little Ice Age (around 1850), global temperatures have begun to increase quickly and steeply (HEGERL *et al* 2018) triggering great changes in the physiognomy of high altitude landscapes (Fig. 1).

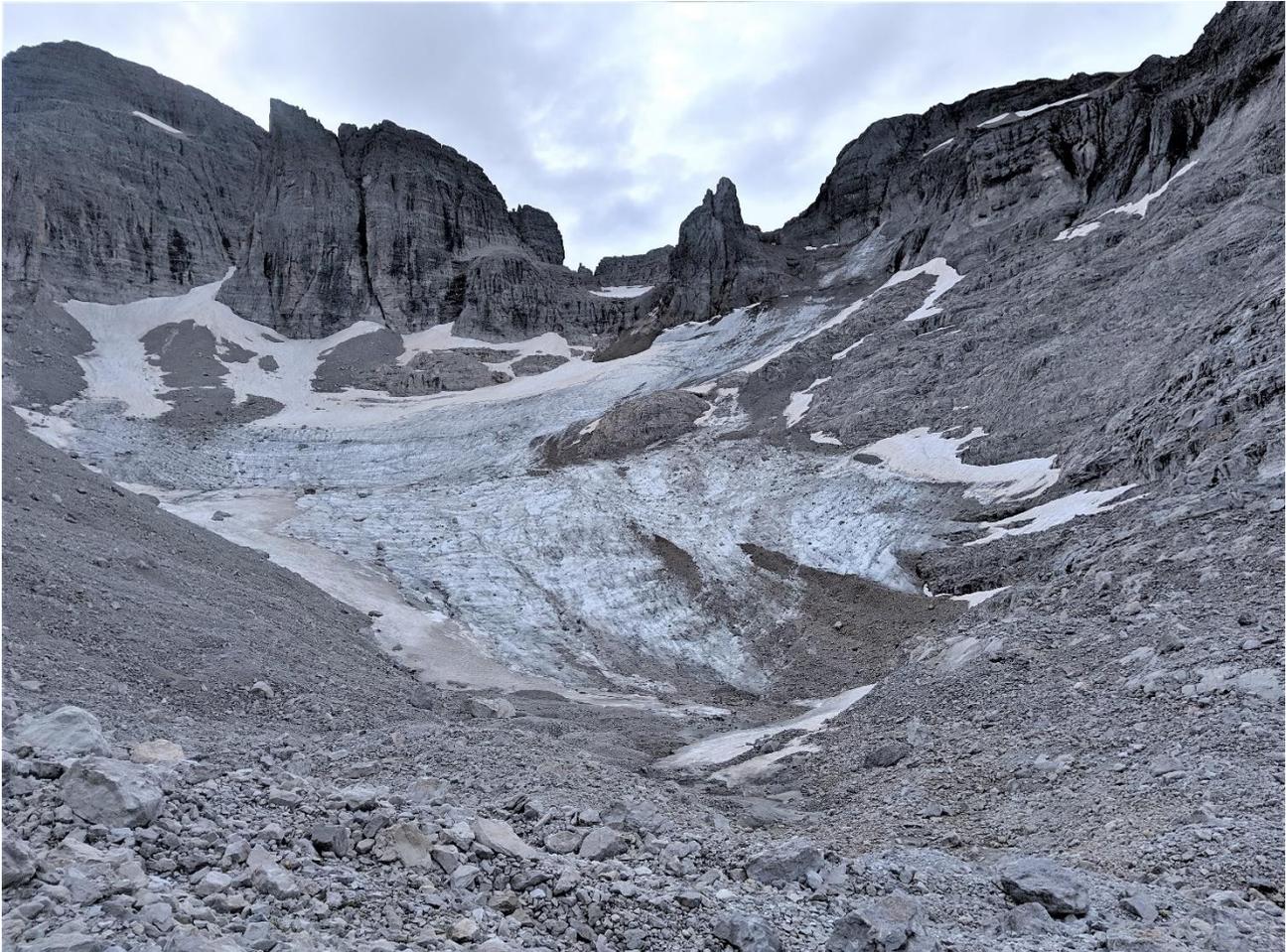

*Figure 1 – Vedretta d'Agola glacier (Brenta Dolomites, Italy); in 2018 (photo by M. Gobbi). This glacier is near to extinction, but still hosts a small population of Nebria germari Heer 1837 on its surface and close to its front (M. Gobbi, unpublished data).*

The current biodiversity loss due to climate change and the destruction or disappearing of habitats is a major societal challenge (ROCKSTRÖM *et al* 2009). However, studies that explored current processes triggered by climate warming on alpine biodiversity are still few.
Understanding the current conservation status of climate-sensitive taxonomic groups is a key starting point for any species monitoring program, and identifying recent trends in abundance and diversity can help in prioritizing focal species and focal landforms for management. Models that relate the distribution of alpine plants and climate to predict the future geographical range of species in response to forecast climate change predict that species living at high altitudes will be strongly affected (STEINBAUER *et al* 2018). Such information is largely lacking for arthropods across much of the alpine areas notwithstanding their biodiversity is higher than vascular plants or vertebrates. Possible reasons for this include i) the lack of historical datasets, ii) the unsatisfactory taxonomical knowledge, or unstable taxonomy, of some relevant taxa, iii) the logistical challenges associated with sampling in inaccessible places and in collecting quantitative data in a standardized way.

The purpose of this paper is to synthesize the response and future perspectives of ground beetles (Coleoptera: Carabidae) to the ongoing changes in the high mountain landscapes of the European Alps through the analysis of three of the most visible abiotic effects of global warming: glacier retreat, increasing stony-debris cover on the glacier surface, and the fusion of the permafrost. Ground beetles are used as flagship taxa because they are sensitive to the effects of climate change, but also because of the availability of two decades of study of high altitude species in the European Alps.

**EFFECTS OF CLIMATE WARMING ON HIGH ALTITUDE ABIOTIC COMPONENT**
Geomorphic processes in high-altitude mountain areas are sensitive to the effects of climate change, which modulate the interaction between glacial and periglacial processes (HARRIS & MURTON 2005). Three of the most visible effects of global warming on the abiotic conditions of high altitude landscapes are i) the glacier retreats, ii) the increasing rock debris cover on glacier surfaces, and iii) the fusion of the permafrost (HARRIS *et al.* 2001; ZEMP *et al* 2009; SCHERLER *et al* 2011). Glaciers are retreating all over the world. Recent data from the Italian Alps show that the observed glacier melt occurred during the last decades was mainly caused by increased ablation, caused by warmer temperatures and related feedbacks, such as the lengthening of the ablation season. The total precipitation does not show any significant trend, but the fraction of solid precipitation decreased as a consequence of the warmer temperature (CARTURAN *et al* 2016).

The increasing rock debris cover on glacier surfaces is causing them to change from clean (or white) glaciers to black ones (also called debris-covered glaciers). The phenomenon can be explained by the progressive exposure of englacial debris to ice melting, and by the increasing occurrence of rock-falls from the slopes freed by glacier thinning and thus exposed to macro-gelivation processes (TAMPUCCI *et al* 2017).

Permafrost is ground, including rock or soil, with a temperature that remains at, or below, 0 °C for two or more years, and is one of the cryospheric indicators of global climate change (HARRIS *et al.* 2001). Most of the permafrost is located at high latitudes (in and around the Arctic and Antarctic regions); at lower latitudes, alpine permafrost occurs at high elevations. Permafrost reacts sensitively to changes in atmospheric temperature, and is warming at global scale (BISKABORN *et al* 2019). The implications of ongoing climate change on cold-adapted species may therefore be significant.

**EFFECTS OF CLIMATE WARMING ON HIGH ALTITUDE BIOTIC COMPONENT: GROUND BEETLES AS FLAGSHIP TAXON**
Ground beetles can be considered among the most important meso- and macro-invertebrates living on recently deglaciated terrain and on glaciers, in terms of both species richness and abundance (GOBBI & LENCIONI, 2020). There is awareness about the extinction risk for some endemic cryophilous species due to habitat destruction (e.g. glacier disappearing) or changes in micro-habitat conditions (e.g. permafrost melt).

Currently, rapid climate warming represents a major concern for high-altitude carabid beetles, specifically for the species living in glacialised mountain areas. Thus, to investigate how high altitude, cold-adapted carabid species are responding to climate change is an urgent goal and an exciting challenge in alpine biogeography and applied entomology.

Most of the carabid species living at high altitudes are endemic and cold-adapted, have low dispersal abilities and form small and/or isolated populations. These traits increase their extinction risk. For instance, an uphill shift and local extinction were demonstrated for some species in the Andes and in the European Alps. MORET *et al* (2016) reported the first evidence for upslope shift of the species *Dyscolus diopsis* (Bates, 1891) along the slopes of the Pichincha volcano in Ecuador. Over one hundred years, this species shifted 300 m upward. Consequently, its distribution area was reduced by >90%, which points to an increased probability of future local extinctions on this volcano as well as on other mountains of Ecuador. A similar altitudinal shift of about 300 m upward, causing local population extinction was reported on the Dolomites (Eastern Italian Alps) for *Nebria germari* Heer, 1837. Until the 1980, this species was common on the high alpine prairies and near glacier fronts (MARCUZZI 1956, BRANDMAYR & ZETTO BRANDMAYR 1988, BERNASCONI *et al* 2019). PIZZOLOTTO *et al* (2014) and BERNASCONI *et al* (2019) provided historical documentation about the area contraction of several populations, the extinction of the species on the alpine prairies, and limiting its presence to high altitude scree slopes, glacier fronts and glacier surfaces.

Ground beetles, similarly to other ectothermic species, develop physiological adaptations in response to changing environments. Beckers *et al* (2020), demonstrate a negative correlation between body size and elevation in *Amara alpina* (Paykull, l790), a typical species of alpine habitats, explaining this as an adaptation to live in harsh environments. Specifically, they argue that high altitude causes lower food availability, shorter season length, lower temperatures, and increased transpiration risk due to wind exposure. Interestingly, these observations agree with those reported by Gobbi *et al* (2010) about *Oreonebria castanea* (Bonelli, 1810), a typical carabid of the alpine and nival belt of the Alps, whose body size decreases with increasing distance from the glacier snout.

Variation in certain life history traits, area contraction, population fragmentation and shift towards high altitude makes ground beetles useful "sentinels" of environmental and climate changes in mountain areas.

*CARABIDS COLONISATION PATTERN ALONG GLACIER FORELANDS*

The current rapid glacier retreat opens up a perfect setting for studying carabids succession under rapidly warming conditions. A glacier foreland can be defined as the area between the current leading edge of the glacier and the moraines of the latest maximum, the Little Ice Age (Fig. 2). When the chronology of glacier retreat is available, we can investigate the spatio-temporal succession and colonisation processes of carabid beetles as well as of other biotic components. Although the investigations on carabid colonization along alpine glacier forelands in the European Alps began in the middle of the last century (JANETSCHEK 1958, DE ZORDO 1979, FOCARILE 1976), the first quantitative study, focusing on the genus *Nebria*, appeared later ( GEREBEN 1995). KAUFMANN (2001) performed the first comprehensive study on invertebrate primary succession along an Alpine glacier foreland, also including data on carabid beetles . Since then, several papers were published on carabid beetle succession along European glacier forelands (e.g. GOBBI *et al* 2007, SCHLEGEL *et al* 2012 in the Alps, VATER & MATTHEWS 2015, HÅGVAR *et al* 2017 in Norway), and more recently on an equatorial glacier foreland (MORET *et al.* 2020).

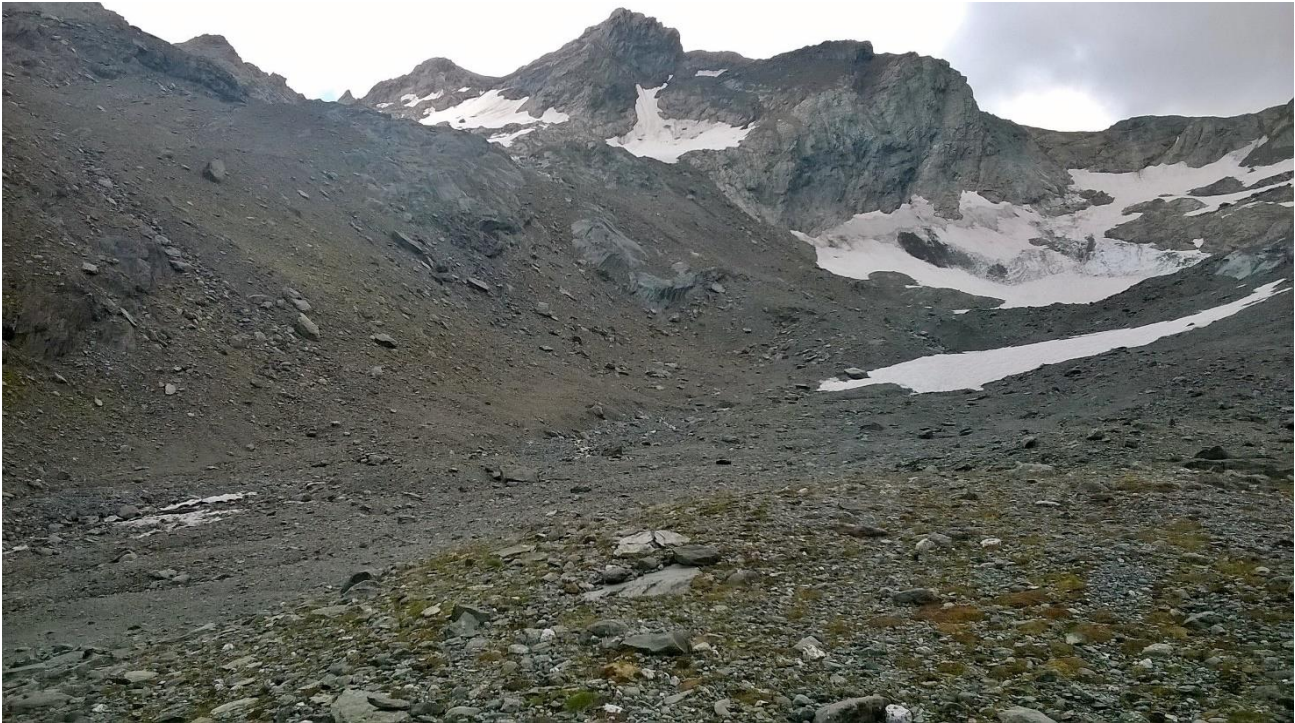
*Figure 2 – Example of a glacier foreland (Trobio glacier, Orobian Alps, Italy; photo by M. Gobbi).*

Time since deglaciation seems to be the main variable affecting carabid beetle species richness, taxonomic and functional diversity patterns in space and time. Local micro-habitat features (e.g. ground temperature and humidity, soil grain size, pH, soil content of organic matter, vegetation cover) are strong drivers of the species distribution and population size. KAUFMANN (2001) described the colonisation pattern as deterministic and directional, but more recently, VATER & MATTHEWS (2015) proposed two different modes of colonisation: the 'addition and persistence' and 'replacement-change'. The former was observed in Northern-Europe, in the Andes (VATER & MATTHEWS 2015; MORET *et al* 2020) and on peripheral mountain groups of the Southern Alps (TAMPUCCI *et al* 2015). It consists in the persistence of pioneer species (i.e. the initial colonisers, e.g. *Nebria* spp.) from the recently deglaciated sites (early successional stages) to sites deglaciated >100 years ago (late successional stages). In this case, there is no species turnover along the chronosequence of glacier retreat. During the 'replacement-change' process, mainly observed in the Alps (e.g. SCHLEGEL *et al* 2012, GOBBI *et al* 2017), a group of initial colonizers (the pioneer community) are progressively replaced over time by one or more other species; in this case there is a clear species turnover.

Due to different biogeographic history, the glacier forelands in Northern Europe show different pioneer species than those located in the European Alps, but they share the same genera. In northern European glacier forelands, the pioneer species (species living on terrains deglaciated for <20 years), include *Amara alpina*, *Amara quenseli*, *Nebria nivalis, N. rufescens, Bembidion hastii* and *B. fellmanni* (BRÅTEN *et al* 2012; VATER & MATTHEWS 2015). In the European Alps, the pioneer community is mainly formed by *Nebria*/*Oreonebria* species, most of them steno-endemic to specific mountain groups or areas of the Western, Central and Eastern Alps (Fig. 3), with the occasional addition of species of Bembidiinae (e.g. *Sinechostictus doderoi*, *Bembidion bipunctatum*) and Zabrini (*Amara quenseli*). In pioneer stages, prey availability can be extremely patchy and episodic.

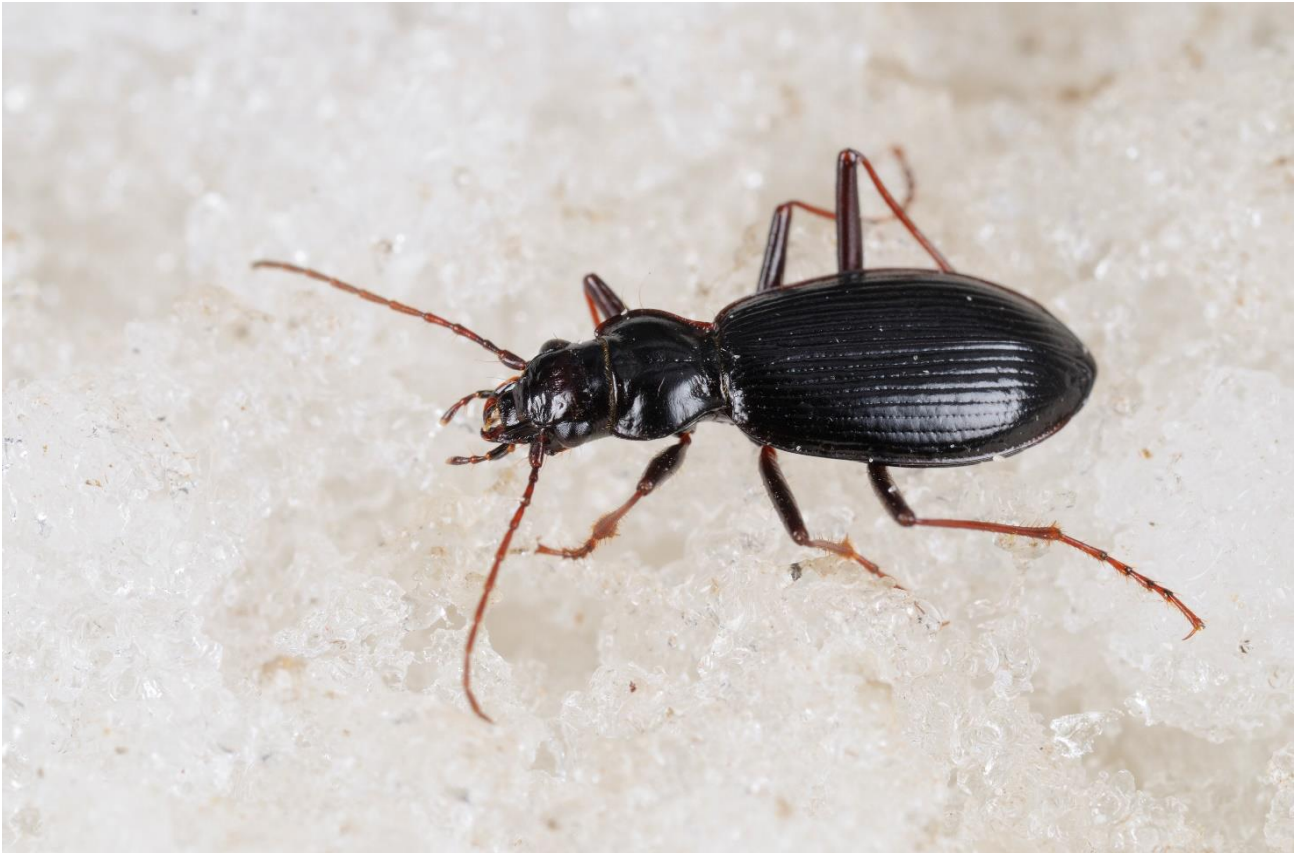

*Figure 3 – Nebria germari Heer 1837 walking on the Presena glacier (Presanella Group, Central-eastern Italian Alps; photo by F.Pupin/archive MUSE).*

Sint *et al* (2018) demonstrated that prey found in carabid guts were mainly collembolans, spiders and allochthonous flying insects (e.g. aphids of the genus *Cinara*).

The colonisation of a glacier foreland proceeds in space and time as the succession proceeds (i.e. vegetation cover). Starting about 20y after deglaciation, to late successional stages (terrains deglaciated since the end of the LIA or more) carabid communities gradually become taxonomically and functionally more distinct. During this period, *Amara* spp. (omnivorous, seed-eaters), *Carabus* spp. and *Leistus* spp. (specialised predators), *Cymindis* and *Patrobus* spp. (predators) gradually appear.

Few data are available about the speed of colonisation. VATER & MATTHEWS (2015) claim that the speed of colonisation is higher along glacier forelands crossing the subalpine zone than those above the treeline. In addition, KAUFMANN (2002) estimated that an increase of 0.6°C in summer temperatures will double the speed of initial colonisation. BRAMBILLA & GOBBI (2014) modelled the time needed by carabids to effectively react to habitat modification and highlighted that sites already hosting the suitable land cover type, but ice-free for < 100y are mainly colonised by winged carabid beetles; a delayed response (ca. 100 yrs) to environmental modifications due to glacier retreat was mostly in brachypterous carabids. This time lag implies caution in predicting species range shifts following climate change.

Therefore, it is evident that if the speed of adaptive capacity or dispersion is not temporally synchronous with the speed of the glacier retreat, the only way cryophilous species can survive is to colonise refuge areas.

*ARE THERE ALPINE LANDFORMS ACTING AS PRESENT CLIMATIC REFUGIA?*

Refugia are places where species can persist when the conditions on the overall landscape become unfavourable, and from where they can potentially expand when environmental conditions change for the better (microrefugial equilibrium, RULL 2009). Traditionally, the current species distribution in climate-limited ecosystems, like those located in mountain regions, have been described as cold-stage refugia (i.e. glacial refugia) during the Last Glacial Maximum (c. 22000 years BP).

The role of present climatic refugia, landforms that support locally favourable climates within larger areas with unfavourably warmer climates (warm-state refugia, Gentili *et al* 2015), is a neglected aspect in the study of carabid responses to climate change (but see KAVANAUGH (1979). KAVANAUGH (1979) hypothesised the presence of mountain landforms that preserve suitable climatic conditions for cold-adapted species (specifically *Nebria* spp.) in spite of the climate warming. Surely, mountain landscapes created opportunities and challenges for carabid species during the Holocene; thus, the identification of such warm-stage refugia is the most exciting challenge of contemporary alpine biogeography. Therefore, the question is: may some alpine landforms promote the long-term survival of cold adapted carabids when the surrounding habitats become climatically unfavourable?

Several authors argued that ice-related landforms, like debris-covered glaciers and rock glaciers, can act as potential refugia for cold-adapted plants, arthropods and small rodents (MILLAR *et al* 2010, 2013; GENTILI *et al* 2015; TAMPUCCI *et al* 2017a, b).

Debris-covered glaciers are glaciers partially or totally covered by a continuous layer of sand, gravel and rock; they are particularly common in the mountains of Asia, but the stone debris is fast increasing also on glaciers located in other areas such as the European Alps (SCHERLER *et al* 2011). While a thin debris layer (< 5 cm, the "critical thickness") promotes further ablation through its thermal conductivity, a thicker debris layer will logarithmically decrease the rate of ice melting as a consequence of thermal insulation provided by debris, allowing glaciers to reduce mass loss (MIHALCEA *et al* 2007) (Fig. 4).

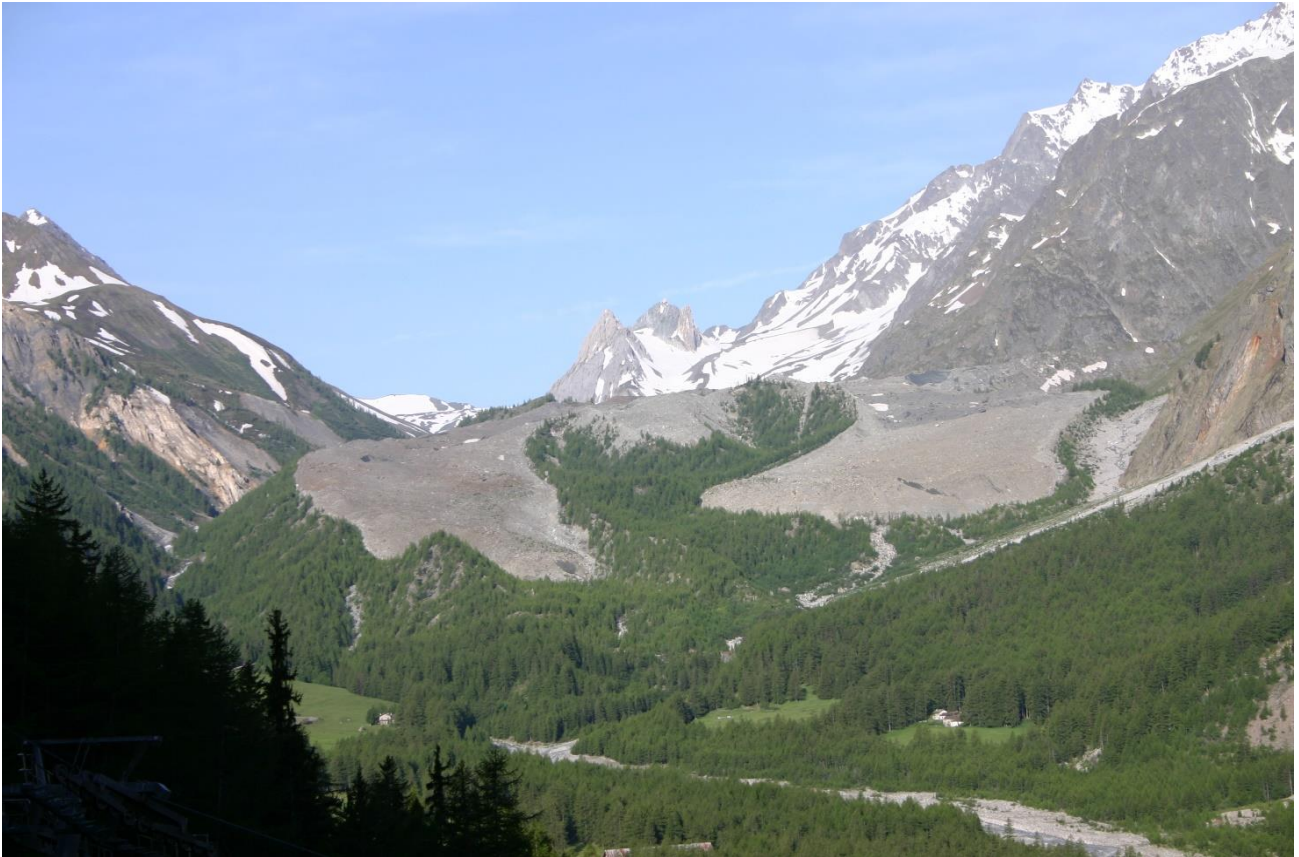

*Figure 4 – Example of a debris-covered glacier descending below the treeline (Miage glacier, year 2006; Mount Blanc Group, Western Italian Alps; photo by M. Gobbi). It hosts permanent populations of Oreonebria angusticollis (Bonelli, 1809) (GOBBI et al 2011).*

Rock glaciers are lobate or tongue-shaped landforms consisting of rock debris with either an ice core or an ice-cemented matrix; they are the most spectacular and common periglacial phenomena on mountains (JANKE 2013). There are different kinds of rock glaciers (e.g. active, inactive and fossil) in relation to their movement and presence of interstitial ice.

The thermal profile and inertia observed on some alpine debris-covered glaciers and active rock glaciers supports the hypothesis that they can act as refuge areas by decoupling the topoclimate from the regional climate. Specifically, they differ from the surrounding landforms (e.g. scree slopes) by an overall lower ground surface temperature (average annual temperatures around or below 0°C) (GOBBI et al 2014, TAMPUCCI et al 2017a, b, GOBBI et al 2018) (Fig. 5).

Permanent populations (adults plus larvae at different stages of development) of carabid beetles on debris-covered glaciers and rock glaciers are documented only in the European Alps.

These ice-related landforms can host populations of *Nebria* spp. (e.g. *Nebria germari* in Eastern Italian Alps; GOBBI et al 2017, BERNASCONI et al 2020) and *Oreonebria* spp. (e.g. *Oreonebria angusticollis* in Western Italian Alps, GOBBI et al 2011; *Oreonebria castanea* Central Italian Alps, GOBBI et al 2007).

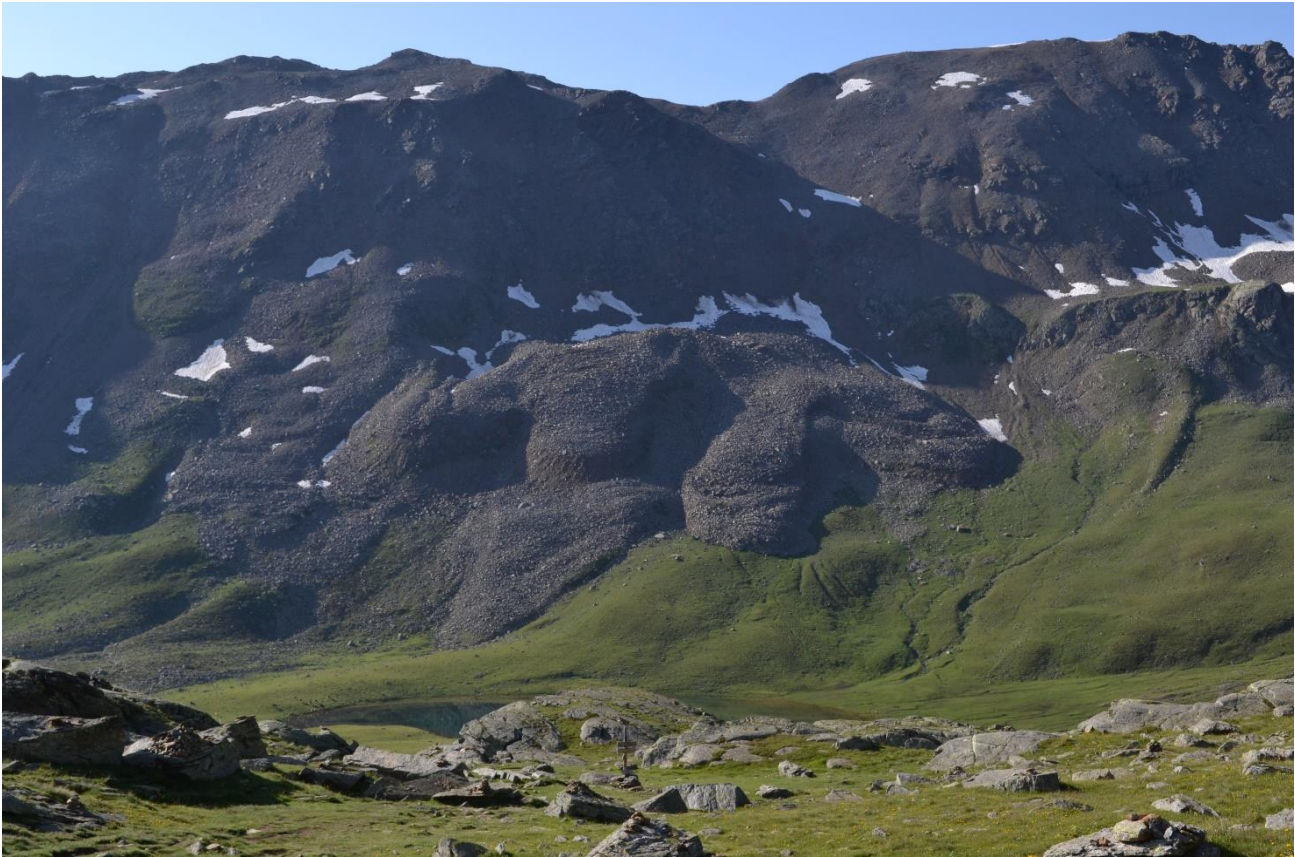

*Figure 5 – Example of rock glacier with its typical lobate tongue (Lago Lungo, Ortles-Cevedale Group, Central-eastern Italian Alps; photo by Duccio Tampucci).*

Specifically, debris-covered glaciers with their tongue descending below the climatic treeline, thus crossing the subalpine belt, permit the survival of cold-adapted carabids below their normal altitudinal distribution as a consequence of the thermal effect of underlying ice which ensure annual temperatures around 0°C (e.g. GOBBI *et al* 2011; TAMPUCCI *et al* 2017a). In addition, the pressure of the stones on the ice located under them ensure the constant presence of a thin layer of water that maintains high levels of relative humidity (above 90%, see GOBBI *et al* 2017 and TAMPUCCI *et al* 2017a) which provides suitable micro-habitat for the larvae (M. GOBBI, pers. obs.). Interestingly, the detection probability of the carabids living on debris-covered glaciers is negatively correlated with the percentage of gravel and the thickness of the stony debris on the glacier (TENAN *et al* 2016). This is because carabids move down in a three-dimensional space searching, in the depth of the stony detritus, favourable micro-climate conditions to escape from unfavourable temperatures near the surface. Temperatures at a depth of 10 cm can reach almost 30°C during the summer (GOBBI *et al* 2017).

The carabid populations observed on the debris-covered glaciers are female-biased; this could be an adaptaion to maintain viable populations in harsh environments (TENAN *et al* 2016).

Data collected on active rock glaciers in the Alps highlighted a negative correlation between their importance as refugia and their altitude. Strong differences in cold-adapted carabid abundance between active rock glacier and surrounding ice-free landforms (scree slopes and alpine prairies) were found on the rock glaciers located at the lowest altitudes (ca. 2400 m asl or below). In these cases, the population size of some cold-adapted species like *Oreonebria castanea*, *O. soror*, *Nebria germari* and *Trechus tristiculis* is at least three times higher than those recorded on neighbour scree slopes (GOBBI *et al* 2014). Currently, in some scree slopes, these cold-adapted species are completely absent due to the uphill shift of species linked to more vegetated, dry and warm habitats (e.g. *Carabus adamellicola, C.sylvestris, C.concolor, Pterostichus multipunctatus, Cymindis vaporariorum, Cymindis cordicollis*) (GOBBI *et al* 2014, TAMPUCCI *et al* 2017b). Conversely, above 2600 m asl, active rock glaciers and scree slopes still support large populations of cold-adapted carabids due to the long-lasting snow cover. For instance, *O.castanea* activity density (no. of individuals/days of trap activity) recorded on the Col d'Olen rock glacier (Rose Massif, Western Italian Alps) was 0.15 on rock glacier and 0.1 both on scree and vegetated (mature) slopes (unpublished data).

To date, we don't know how long this warm stage period persists and whether these ice-related landforms will act as refugia during the entire warm period, but there is clear evidence of their current role in mitigating and slowing down the extinction risk of cold-adapted carabid species.

## *THE NEED FOR AN HIGH-ALPINE BIODIVERSITY MONITORING PROGRAMME*

Challenge for the future is to increase the number of baseline studies to summarise the status and trends of high altitude and cold-adapted species. a useful framework for this could be a High-alpine Biodiversity Monitoring Programme similar to the ongoing Circumpolar Biodiversity Monitoring Programme (CHRISTENSEN *et al* 2013).

It would be clearly important to develop a sampling protocol that is as easy as possible to carry out, both in terms of time-effectiveness and labor intensity due to the constraints imposed by the physical effort of working at high altitudes: walk for several hours off the beaten paths, weather instability, and a heavy backpack.

On the base of the available methodological studies performed at high altitude (HARRY *et al* 2011; GOBBI *et al* 2018) the following suggestions can be provided:

i) an early activity peak is predominant among carabids in mountain ecosystems, thus the sampling season should starts just after the snow melts (generally early July),
ii) two fortnightly sampling sessions are recommended to obtain a species inventory, while three to five sampling sessions are suggested to obtain data on population dynamics and population size (activity density),
iii) three to six traps, located about 10 m apart, are needed for each sampling plot (= area with uniform environmental features),
iv) use plastic pots (diameter 7 cm, height 10 cm) that are smaller than those commonly used (diameter 9 cm, height 11cm; 150 ml of preservative solution). A smaller trap permits the use of less preservative (ca. 50 ml), so it is possible to put on the ground a higher number of traps, thus sampling points (it means more robust analysis) (Fig. 6),
v) bait the vessel with a solution of wine-vinegar, salt and a drop of soap, this mixture is not toxic thus freely usable in protected areas (e.g. Natural and National Parks),
vi) perform the statistical analysis considering each trap separately (pseudoreplicate) and using spatially explicit models which are able to handle spatially autocorrelated data (GOBBI & BRAMBILLA 2016).

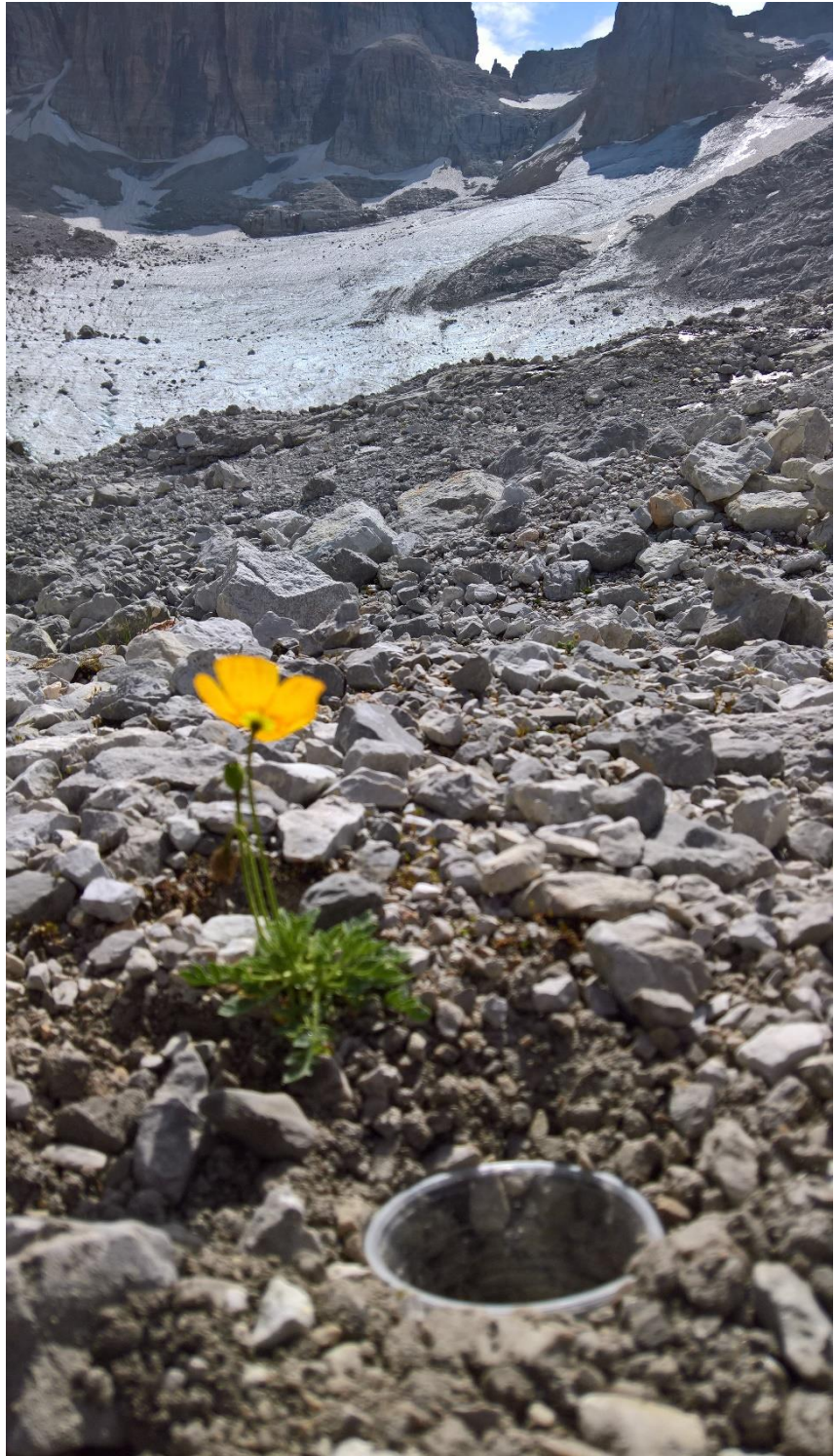

*Figure 6 – Pitfall trap placed along the Vedretta d'Agola glacier foreland (Brenta Dolomites, Italy; photo by M. Gobbi).*

**CONCLUSIONS**

This paper provides the first comprehensive synthesis of the knowledge about carabid beetles in ice-related high altitude alpine landforms.

Summarising, what we currently know is that i) past climate events shaped the current distribution of several species, ii) cold-adapted carabids are negatively affected by climate warming and iii) extinction and upslope shift have been already documented for several species.

What we should know is that current climate events are shaping the distribution of several species, and some high altitude ice-related landforms like debris-covered glaciers and rock glaciers could act as warm-stage refugia thanks to their thermal inertia.

Recognized potential refuge areas and monitor size and dynamic of the hosted carabid populations is extremely important to obtain detailed information about the effect triggered by climate warming on high altitude biodiversity.

The physiognomy of high altitude landscapes is quickly changing, specifically in areas covered by glaciers. Carabid beetles are good candidates as bioindicators of these changes. There are many endemic and micro-endemic species that we are risking to lose, and this calls for long term and standardised monitoring programs.

**Acknowledgements -** I am grateful to the Organizing Committee of the XIX European Carabidologists Meeting held in Paneveggio-Pale di San Martino Natural Park (Italy) for inviting me as keynote speaker. I also thank Roberto Pizzolotto, Gabor Lövei and an anonymous referee for their suggestions on the manuscript and Pietro Brandmayr, David Kavanaugh and Pierre Moret who triggered my interest in studying high altitude carabid beetles. Last, but not least, thanks to my colleagues and students that since more than 15 years are cooperating with me in collecting carabids in harsh environments of the world, as well as assist in analyzing the spatial and temporal patterns.